# Interference Alignment with Limited Feedback on Two-cell Interfering Two-User MIMO-MAC

Namyoon Lee, Wonjae Shin and Bruno Clerckx
Samsung Advanced Institute of Technology (SAIT), Yongin-si, Korea
Email: namyoon.lee@gmail.com, {wonjae.shin, and bruno.clerckx}@samsung.com

*Abstract*— In this paper, we consider a two-cell interfering two-user multiple-input multiple-output multiple access channel (MIMO-MAC) with limited feedback. We first investigate the multiplexing gain of such channel when users have perfect channel state information at transmitter (CSIT) by exploiting an interference alignment scheme. In addition, we propose a feedback framework for the interference alignment in the limited feedback system. On the basis of the proposed feedback framework, we analyze the rate gap loss and it is shown that in order to keep the same multiplexing gain with the case of perfect CSIT, the number of feedback bits per receiver scales as $B \geq (M-1)\log_2(\mathsf{SNR}) + C$, where $M$ and $C$ denote the number of transmit antennas and a constant, respectively. Throughout the simulation results, it is shown that the sum-rate performance coincides with the derived results.

## I. INTRODUCTION

Communications over wireless medium suffer from the interference since the broadcast characteristic of the wireless medium naturally causes interference among communication nodes. Therefore, interference management has played an important role in developing encoding and decoding schemes to achieve channel capacity in various wireless networks. Recently, by introducing interference alignment, the authors in [1]-[3] provided a way of successfully resolving the interference problem in wireless networks. The basic concept of interference alignment is to put all interference signals in small signal dimensions at each receiver so as to minimize the signal dimensions occupied by the interference signals while independently keeping the desired signals dimension. Using this scheme in the multiple antenna system, many researchers have investigated the multiplexing gains of a variety of wireless networks such as $X$ networks [1], [2], interference network [5], [6], compound broadcast channel [7], interfering broadcast channel [8], [9], and a multiuser bi-directional relay network ($Y$ channel [10]).

However, most of the previous studies relied on the assumption of complete channel state information at the transmitter (CSIT) for all users. In practice, because this assumption is highly unrealistic, transmit strategies under the assumption of channel uncertainty at the transmitter have been studied [11]-[14]. By considering interference alignment with limited feedback, the authors in [11] showed that $K/2$ multiplexing gains can be achieved for $K$-user single-input-single-output (SISO) frequency selective interference channel with $L$ taps if each receiver feedbacks the total number of feedback bits at least $K(L-1)\log(\mathsf{SNR})$ to all transmitters and receivers except itself. By applying the multiple antennas system, the authors [12] extended the result of [11]. In [12], they demonstrated that the same multiplexing gains as the original result in [4] is achieved as long as each receiver exploits no less than $\min\{M,N\}^2 K(RL-1)\log(\mathsf{SNR})$ bits feedback rate for $K$-user MIMO channel with $M$ antennas at each transmitter and $N$ antennas at each receiver, where $R = \lfloor \frac{\max\{M,N\}}{\min\{M,N\}} \rfloor$. In a cellular system, under the assumption of channels estimation errors, the authors in [13] investigated the achievable sum-rate using the interference alignment scheme. Furthermore, the author in [14] proposed a simple interference alignment scheme employing the knowledge of channel structure such as correlations instead of using explicit CSIT.

In this paper, we consider a two-cell interfering two-user multiple-input multiple-output multiple access channel (MIMO-MAC) with finite rate feedback, which is the well-matched model with the multi-cell multi-user uplink scenario. Our contributions are summarized as follows:

1) We investigate the multiplexing gain for the two-cell interfering two-user MIMO-MAC under the assumption with perfect channel state information at transmitters (CSIT) using an interference alignment precoder and zero-forcing decoder.

2) In the two-cell interfering two-user MIMO-MAC, we come up with a framework for the interference alignment with finite rate feedback, which employs quantized transmit beamforming vector feedback for the interference alignment instead of quantized channel vector feedback [11], [12].

3) Based on the proposed feedback framework, we also derive the rate gap loss for that system with limited feedback using the random vector quantization (RVQ) method in [16]-[18]. Using this rate gap loss analysis, it is shown that in order to maintain the same multiplexing gain with the case of perfect CSIT, the number of feedback bits per user scales as $B \geq (M-1)\log_2(\mathsf{SNR}) + C$, where $M$ and $C$ denote the number of transmit antennas and a constant, respectively. Interestingly, we show that the proposed feedback framework can significantly reduce the number of feedback bits to maintain the optimal multiplexing gains when the interference alignment is applied under finite rate feedback system compared with the previous result in [12]. From this, we can see the benefit of the transmit beamforming vector feedback mechanism when the interference alignment is utilized. Throughout the simulation results, it is shown that the sum-rate performance coincides with the derived results.

The remainder of this paper is organized as follows. Section

II describes the signal model used in the current study. We derive the multiplexing gain of the two-cell interfering two-user MIMO-MAC in Section III. In Section IV, we come up with a framework for the interference alignment scheme with limited feedback. In Section V, we analyze for the rate loss due to finite rate feedback and show numerical results to demonstrate the performance of the proposed scheme. Section VI contains the conclusions.

## II. SYSTEM MODEL

In this section, we describe the system model for a two-cell interfering MIMO-MAC with finite rate feedback as shown in Fig. 1. The channel basically consists of 4-users with $M$ antennas and two base stations (BSs) being with $N$ antennas. As shown in Fig. 1, user 1 and user 2 want to convey messages $W_1$ and $W_2$ to $BS_1$, while user 3 and user 4 send messages $W_3$ and $W_4$ to $BS_2$, respectively. The transmit signal at user $i$ is written as

$$\mathbf{x}_i = \mathbf{v}_i s_i, \quad i \in \{1, 2, 3, 4\}, \tag{1}$$

where $s_i$ denotes a symbol for carrying message $W_i$, and $\mathbf{v}_i$ is the linear precoding vector for the symbol. We assume that each user has uplink transmit power constraint, i.e., $\mathbb{E}\left[\text{tr}\left(\mathbf{x}_i \mathbf{x}_i^H\right)\right] \leq P$. The received signals at each BS are represented as

$$\mathbf{y}_j = \sum_{i=1}^{4} \mathbf{H}_{j,i} \mathbf{x}_i + \mathbf{n}_j, \quad j \in \{1, 2\}, \tag{2}$$

where $\mathbf{n}_j$ is $N \times 1$ additive white Gaussian noise (AWGN) vector at $BS_j$, and $\mathbf{H}_{j,i}$ is channel matrix with size of $N \times M$ from user $i$ to $BS_j$. Each entry of the matrices $\mathbf{H}_{j,i}$ for $\forall i, j$ is drawn from independently and identically distributed (i.i.d.) random variable according to $\mathcal{CN}(0,1)$. This ensures that all channel matrices almost surely have full rank, i.e., $\text{rank}(\mathbf{H}_{j,i}) = \min\{M, N\}$. It is assumed that the channel state information (CSI) is available at receiver.

Each BS decodes the desired messages coming from its serving users. The decoded signal for user $i$ is represented as

$$\tilde{y}_i = \mathbf{w}_i^H \left[\sum_{m=1}^{4} \mathbf{H}_{j,m} \mathbf{x}_m + \mathbf{n}_j\right], \tag{3}$$

where $\mathbf{w}_i^H$ denotes the receive beamforming vector for decoding message $W_i$ with unit norm, i.e., $\|\mathbf{w}_i\|_2 = 1$.

The achievable rate for message $W_i$ is given by

$$R_i = \log_2\left(1 + \frac{P\left|\mathbf{w}_i^H \mathbf{H}_{k,i} \mathbf{v}_i\right|^2}{\sigma^2 + P\left|\mathbf{w}_i^H \sum_{m \neq i}^{4} \mathbf{H}_{k,m} \mathbf{v}_m\right|^2}\right), \tag{4}$$

where $k = 1$ if $i \in \{1, 2\}$, and $k = 2$ if $i \in \{3, 4\}$.

### A. Sum multiplexing gain

Multiplexing gain is an important metric for assessing the performance of the signaling in the multiple antenna system in the high SNR regime, which is defined as

$$d_{sum} \triangleq \lim_{\mathsf{SNR} \to \infty} \frac{R_{sum}(\mathsf{SNR})}{\log(\mathsf{SNR})}, \tag{5}$$

where $R_{sum}(\mathsf{SNR}) = \sum_{j=1}^{4} R_j(\mathsf{SNR})$ denotes the network sum rate at signal-to-noise ratio ($\mathsf{SNR} = P/\sigma^2$).

## III. UPLINK IA WITH PERFECT FEEDBACK

In this section, using the interference alignment scheme, we investigate the multiplexing gain of the two-cell interfering two-user MIMO-MAC under the assumption of complete CSIT. For simplicity, we consider a case where each user has $M = 2$ antennas, and each BS is equipped with $N = 3$ antennas throughout the remainder of this paper[1]. The following theorem is the main result of this section.

**Theorem 1**: *In the two-cell two-user MIMO-MAC, the maximum sum multiplexing gain is 4 when $M = 2$ and $N = 3$, i.e,*

$$d_{sum} = \sum_{i=1}^{4} d_i = 4. \tag{6}$$

*1) Converse:* The converse is simply verified using the result in [15]. The two-cell interfering two-user MIMO-MAC is equivalently modeled by the two-user MIMO interference channel (MIMO-IC) in [15] if it is allowed two users in the same cell to perfectly cooperate. Since the cooperation between two users in the same cell does not degrade the multiplexing gain of that channel, we can argue that the maximum multiplexing gain for the two-cell interfering two-user MIMO-MAC is upper-bounded by the multiplexing gain of two-user MIMO-IC, which is $\min\{4M, 2N, \max\{2M, N\}\}$. Accordingly, we can conclude the multiplexing gain of 4 is the maximum multiplexing gain when $M = 2$ and $N = 3$.

*2) Achievability:* The achievability of the **Theorem 1** is provided using a simple interference alignment precoder and zero-forcing decoder. First, let us consider $BS_1$. Since $BS_1$ has $N = 3$ dimensional receive signal space, two ICI signal vectors, $\mathbf{H}_{1,3}\mathbf{v}_3$ and $\mathbf{H}_{1,4}\mathbf{v}_4$ should be aligned within one dimensional signal space so as to decode two desired messages $W_1$ and $W_2$. These ICI alignment conditions can be represented as

$$\text{span}(\mathbf{h}_1^{\mathsf{ICI}}) = \text{span}\left([\mathbf{H}_{1,3}\mathbf{v}_3 \quad \mathbf{H}_{1,4}\mathbf{v}_4]\right) \tag{7}$$

where $\text{span}(\cdot)$ denotes the space spanned by the column vectors of a matrix and $\mathbf{h}_1^{\mathsf{ICI}}$ denotes the intersection subspace of $\mathbf{H}_{1,3}$ and $\mathbf{H}_{1,4}$. By using the Lemma 1 in [10], we can find out the intersection subspace satisfying the condition (7) by solving the following matrix equation,

$$\begin{bmatrix} \mathbf{I}_M & -\mathbf{H}_{1,3} & \mathbf{0} \\ \mathbf{I}_M & \mathbf{0} & -\mathbf{H}_{1,4} \end{bmatrix} \begin{bmatrix} \mathbf{h}_1^{\mathsf{ICI}} \\ \tilde{\mathbf{v}}_3 \\ \tilde{\mathbf{v}}_4 \end{bmatrix} = \mathbf{M}_1 \tilde{\mathbf{v}} = \mathbf{0}. \quad (8)$$

---
[1]The generalized antenna configuration will be included in the journal version of this paper due to space limitation.

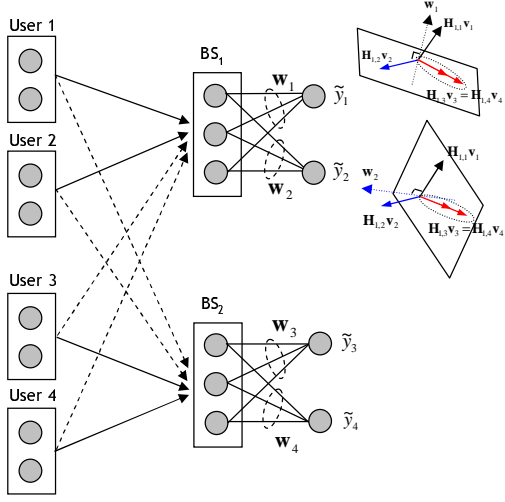

Fig. 1. The system model of the two-cell interfering two-user MIMO-MAC.

Since $\mathbf{H}_{1,3}$ and $\mathbf{H}_{1,4}$ are full-rank and the size of the matrix $\mathbf{M}_1$ is $6 \times 7$, the transmit beamforming vectors, $\mathbf{v}_3$ and $\mathbf{v}_4$, for ICI channel alignment can be obtained with probability one, which are

$$\mathbf{v}_3 = \tilde{\mathbf{v}}_3 / \|\tilde{\mathbf{v}}_3\|, \qquad \mathbf{v}_4 = \tilde{\mathbf{v}}_4 / \|\tilde{\mathbf{v}}_4\|. \quad (9)$$

Accordingly, $BS_1$ can decode $W_1$ and $W_2$ using zero-forcing receive beamforming vectors $\mathbf{w}_1$ and $\mathbf{w}_2$, which are

$$\mathbf{w}_1^H \left[ \mathbf{H}_{1,2}\mathbf{v}_2, \mathbf{h}_1^{\mathsf{ICI}} \right] = \mathbf{0}_{1 \times 2}, \quad \mathbf{w}_2^H \left[ \mathbf{H}_{1,1}\mathbf{v}_1, \mathbf{h}_1^{\mathsf{ICI}} \right] = \mathbf{0}_{1 \times 2}. \quad (10)$$

In the similar way, two ICI signal vectors, $\mathbf{H}_{2,1}\mathbf{v}_1$ and $\mathbf{H}_{2,2}\mathbf{v}_2$, can be aligned within one dimensional intersection subspace, $\mathbf{h}_2^{\mathsf{ICI}}$, at the receiver of $BS_2$ by cooperatively designing $\mathbf{v}_1$ and $\mathbf{v}_2$. Applying the zero-forcing decoders, $BS_2$ can reliably decode messages $W_3$ and $W_4$. As a result, we can achieve the multiplexing gain of 4. ∎

## IV. UPLINK IA WITH LIMITED FEEDBACK

In the previous section, we assume that each user can have perfect CSIT when they design their transmit beamforming vectors. In practice, however, each user can obtain CSI through a limited rate feedback. Therefore, to make the interference alignment scheme feasible we consider the interference alignment scheme under the assumption of the finite rate feedback system. To be specific, we propose the system framework for the interference alignment with limited feedback. The proposed framework involves three steps : 1) Design of beamforming vectors, 2) Quantization of beamforming vectors, and 3) Feedback.

### A. System framework

*1) Design of IA beamforming vectors :* Each $BS_i$ for $i = \{1, 2\}$ obtains CSI for all received channel links, i.e., $\mathbf{H}_{i,j}$. Using these CSI, each BS constucts the transmit beamforming vectors for the other cell users so that ICI signal vectors are aligned in the same signal dimension. For example, $BS_1$ calculates the beamforming vectors $\mathbf{v}_3$ and $\mathbf{v}_4$ so that two ICI signal vectors are aligned in $\mathbf{h}_1^{\mathsf{ICI}}$ as in (7). $BS_2$ also makes the other cell users' beamforming vectors $\mathbf{v}_1$ and $\mathbf{v}_2$ in the same method.

*2) Quantization of beamforming vectors:* Each BS exchanges the beamforming vectors for the other cell users by exploiting error and delay free backhaul channel. After exchanging, each BS quantizes the transmit beamforming vectors for the served users by employing a quantization codebook, $\mathcal{C} = \{\mathbf{c}_l, \mathbf{c}_2, \ldots, \mathbf{c}_Q\}$, each of which consists of $Q$-dimensional unit norm vectors of size $Q = 2^B$, where $B$ is the number of bits for feedback channel. By using minimum chordal distance metric, indices for transmit beamforming vectors are obtained as

$$\hat{\mathbf{v}}_i = \mathbf{c}_{n_i}, \quad n = \arg \max_{1 \leq m \leq 2^B} |\mathbf{c}_m{}^H \mathbf{v}_i|, \quad i = \{1, 2, 3, 4\}. \quad (11)$$

For simple analysis, we apply RVQ model in [16] when we quantize the beamforming vectors.

*3) Feedback:* After quantization, each BS informs the indices $n_i$ to its served users through the limited feedback channel $B$ bps rates. In addition, it is assumed that each BS perfectly feeds back the CQI for channel links to the serving users. Here, CQI is considered as the norm of effective channel vectors. For example, CQI for user 1 is $\|\mathbf{H}_{1,1}\hat{\mathbf{v}}_1\|$.

## V. ANALYSIS

In this section, the rate gap loss compared to perfect CSIT is characterized to better understand the proposed limited feedback framework.

### A. Achievable rate with limited feedback

Let us consider the achievable rate of user 1, i.e., $R_1$. Note that all BSs have the knowledge of all transmit beamforming vectors, i.e., $\mathbf{v}_i$ through the exchange. In order for $BS_1$ to decode the user 1's message, we exploit the receive beamforming vector $\mathbf{w}_1$ as

$$\mathbf{w}_1^H \left[ \mathbf{H}_{1,2}\mathbf{v}_2, \mathbf{h}_1^{\mathsf{ICI}} \right] = \mathbf{0}_{1 \times 2}. \quad (12)$$

As a result, the achievable rate of user 1 with finite rate feedback is given by

$$\begin{aligned} R_1^{LFB} &= \log_2 \left( 1 + \frac{\mathsf{SNR}|\mathbf{w}_1^H \mathbf{H}_{1,1}\hat{\mathbf{v}}_1|^2}{\mathsf{SNR} \sum_{i=2}^4 |\mathbf{w}_i^H \mathbf{H}_{1,i}\hat{\mathbf{v}}_i|^2 + 1} \right), \\ &= \log_2 \left( 1 + \frac{\mathsf{SNR}|\mathbf{g}_{1,1}^H \hat{\mathbf{v}}_1|^2}{\mathsf{SNR} \sum_{i=2}^4 |\mathbf{g}_{1,i}^H \hat{\mathbf{v}}_i|^2 + 1} \right), \quad (13) \end{aligned}$$

where $\mathbf{g}_{1,i}^H = \mathbf{w}_i^H \mathbf{H}_{1,i}$ is the effective channel after applying receive combining vector.

## B. Rate loss relative to perfect feedback

In order to figure out the effect of limited feedback, we characterize the expectation of rate loss per user. The rate loss of user $i$, $\triangle R_i$, is defined as the difference between the achievable rate of the user $i$ under the condition of perfect CSI feedback and limited transmit beamforming vector feedback:

$$\triangle R_i \triangleq \mathbb{E}\left[R_i^{PFB} - R_i^{LFB}\right], \quad (14)$$

where $R_i^{PFB}$ represents the rate achieved for user $i$ by interference alignment with perfect CSI, which is defined as

$$R_i^{PFB} = \log_2\left(1 + \mathsf{SNR}|\mathbf{w}_i^H \mathbf{H}_{l,i} \mathbf{v}_i|^2\right). \quad (15)$$

The following theorem is the main result of this section.

**Theorem 2**: *In the two-cell interfering two-user MIMO-MAC, when interference alignment is applied, the rate loss per user due to finite rate feedback is*

$$\triangle R_i \leq \log_2\left(1 + \mathsf{SNR}\sum_{j=1,j\neq i}^{4} A_j 2^{\frac{-B}{M-1}}\right), \quad (16)$$

Proof:) Consider user 1. By definition of the rate loss in (14), we can rewrite the rate loss of user 1 due to quantization error as in (17) (See the top of the next page). Since log function in (17) is a monotonically increasing function and $\sum_{i=2}^{4}|\mathbf{g}_{1,i}^H \hat{\mathbf{v}}_i|^2 \geq 0$, the rate loss of user 1 is bounded as

$$\triangle R_1 \leq \mathbb{E}\left[\log_2\left(1 + \mathsf{SNR}|\mathbf{g}_{1,1}^H \mathbf{v}_1|^2\right)\right] -$$
$$\mathbb{E}\left[\log_2\left(1 + \mathsf{SNR}|\mathbf{g}_{1,1}^H \hat{\mathbf{v}}_1|^2\right)\right] + \mathbb{E}\left[\log_2\left(\mathsf{SNR}\sum_{i=2}^{4}|\mathbf{g}_{1,i}^H \hat{\mathbf{v}}_i|^2 + 1\right)\right]$$
$$= \mathbb{E}\left[\log_2\left(\mathsf{SNR}\sum_{i=2}^{4}|\mathbf{g}_{1,i}^H \hat{\mathbf{v}}_i|^2 + 1\right)\right]. \quad (18)$$

In (18), the last equality comes from the fact that $\mathbf{v}_1$ and $\hat{\mathbf{v}}_1$ are designed independently with respect to $\mathbf{g}_{1,1}^H$ and isotropically distributed in $\mathbb{C}^M$ as shown in both lemma 1 and lemma 4. By applying Jensen's inequality to the last term in (18), the upper bound of the rate loss for user 1 is given by

$$\triangle R_1 \leq \log_2\left(\mathbb{E}\left[\mathsf{SNR}\sum_{i=2}^{4}|\mathbf{g}_{1,i}^H \hat{\mathbf{v}}_i|^2\right] + 1\right). \quad (19)$$

To see the impact on finite rate feedback more specifically, we decompose the quantized transmit beamforming vector, i.e., $\hat{\mathbf{v}}_j$, into two orthogonal basis by using real transmit beamforming vector and residual error vector as

$$\hat{\mathbf{v}}_j = \mathbf{v}_j(\cos\theta_j) + \mathbf{e}_j(\sin\theta_j), \quad (20)$$

where $\theta_j$ stands for the angle between the quantized transmit beamforming direction and real beamforming vector direction for user $j$. In addition, $\mathbf{e}_j$ denotes the error vector because of quantization process. As a result, the rate loss of user 1 in (19) is rewritten as

$$\triangle R_1 \leq \log_2\left(\mathsf{SNR}\sum_{i=2}^{4}\mathbb{E}\left[\|\mathbf{g}_{1,i}\|^2 \sin^2\theta_i \|\bar{\mathbf{g}}_{1,i}^H \mathbf{e}_i\|^2\right] + 1\right), \quad (21)$$

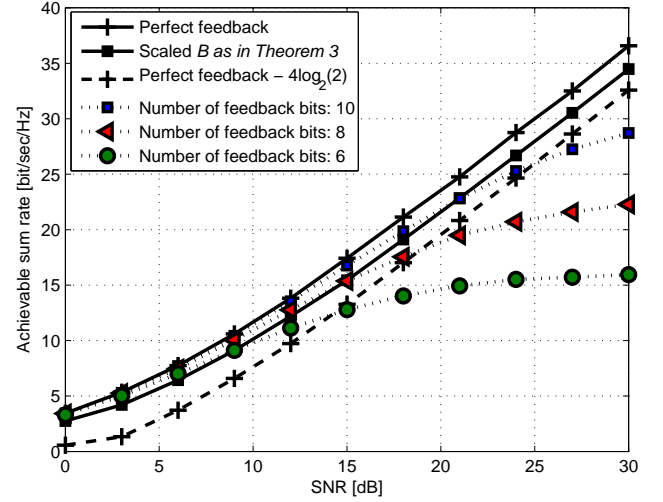

Fig. 2. Sum-rate performance of the proposed feedback scheme.

where $\bar{\mathbf{g}}_{1,i} = \frac{\mathbf{g}_{1,i}}{\|\mathbf{g}_{1,i}\|}$ is the effective channel direction vector. Note that, from lemma 4 in the Appendix, $|\bar{\mathbf{g}}_{1,i}^H \mathbf{e}_i|^2$ are Beta-distributed random variables with parameters $(1, M-2)$. Furthermore, by using the fact that random variables $\|\mathbf{g}_{1,i}\|^2$, $\sin^2\theta_i$, and $|\bar{\mathbf{g}}_{1,i}^H \mathbf{e}_i|^2$ are linearly independent, we finally obtain the upper bound of the rate loss as

$$\triangle R_1 \leq \log_2\left(1 + \mathsf{SNR}\sum_{i=2}^{4} A_i 2^{\frac{-B}{M-1}}\right), \quad (22)$$

where we use the fact in lemma 2 in the Appendix, which gives $\mathbb{E}(\sin^2\theta_i) \leq 2^{\frac{-B}{M-1}}$ and $A_i$ denotes the expectation of the squared norm of effective interference channel ,i.e, $\mathbb{E}\left[\|\mathbf{g}_{1,i}\|^2\right]$ [2]. In a similar way, the rate lose for user $i$, i.e, $\triangle R_i$, is the same with the result in (16). ∎

## C. Scaling law of feedback bits per user

Using **Theorem 2**, we now derive the scaling law of feedback bits per user for obtaining the same multiplexing gain with perfect CSIT case in **Theorem 1**.

**Theorem 3**: *In the two-cell interfering two-user MIMO-MAC, to keep a constant rate loss of $\log_2 \tau$ bps/Hz compared to perfect CSIT, the number of feedback bits for each user scales as*

$$B \geq (M-1)\log_2(\mathsf{SNR}) + C, \quad (23)$$

*where $C$ denotes a constant, which is defined as $C = (M-1)\log_2\left(\frac{\sum_{j=1,j\neq i}^{4} A_j}{\tau - 1}\right)$.*

---

[2]Even if we do not derive the exact distribution of $\|\mathbf{g}_{1,i}\|^2$, its expectation can be considered as a constant term which has impact on increasing the rate gap loss. In fact, we check that $\|\mathbf{g}_{1,i}\|^2$ for $i = 2, 3, 4$ is gamma distributed with parameters $(1.16, 1.3)$ through the maximum likelihood estimates (MLE) statistic test when $M = 2$ and $N = 3$. Thus, we can see $\mathbb{E}\left[\|\mathbf{g}_{1,i}\|^2\right] \simeq 1.5$ for $i = 2, 3, 4$. The rigorous proof for this distribution will be included in the journal version of this paper.

$$\triangle R_1 = \mathbb{E}\left[\log_2\left(1+\mathsf{SNR}|\mathbf{g}_{1,1}^H\mathbf{v}_1|^2\right)\right] - \mathbb{E}\left[\log_2\left(1+\frac{\mathsf{SNR}|\mathbf{g}_{1,1}^H\hat{\mathbf{v}}_1|^2}{\mathsf{SNR}\sum_{i=3}^{4}|\mathbf{g}_{1,i}^H\hat{\mathbf{v}}_i|^2+1}\right)\right]$$

$$= \mathbb{E}\left[\log_2\left(1+\mathsf{SNR}|\mathbf{g}_{1,1}^H\mathbf{v}_1|^2\right)\right] + \mathbb{E}\left[\log_2\left(\mathsf{SNR}\sum_{i=2}^{4}|\mathbf{g}_{1,i}^H\hat{\mathbf{v}}_i|^2+1\right)\right] - \mathbb{E}\left[\log_2\left(1+\mathsf{SNR}\sum_{i=1}^{4}|\mathbf{g}_{1,i}^H\hat{\mathbf{v}}_i|^2\right)\right]. \quad (17)$$

Proof:) To be within $\log_2(\tau)$ (bps/Hz), the rate loss can be represented as

$$\log_2 \tau \geq \log_2\left(1+\mathsf{SNR}\sum_{j=1,j\neq i}^{4} A_j 2^{\frac{-B}{M-1}}\right). \quad (24)$$

If we express equation (24) in terms of feedback bits per user, $B$, then we obtain the inequality condition in (23). ■

*Remark 1.* The derived feedback bits requirement in **Theorem 3** is the same result with the multi-user MIMO broadcast system in [17]. However, it is quite small amounts compared with the required feedback bits for $K$-user MIMO interference channel with limited feedback [12]. Therefore, in order to apply the interference alignment in the limited feedback system with a reasonable amount of feedback bits, the quantized transmit beamforming vector feedback method is more efficient than the quantized channel feedback scheme [12].

### D. Simulation results

In this simulation, we verify the result of **Theorem 3**. Throughout the simulation, we consider parameters, $M = 2$, $N = 3$, $\tau = 2$, and $A_i = 1.5$ for all $i$. As shown in Fig. 2, we can see that the interference alignment using the proposed feedback method maintains the sum of rate loss within $4\log_2(2) = 4$ (bps/Hz) with respect to the case with perfect CSIT by increasing $B$ as in (23). However, if we fix the number of feedback bits, the sum rate performances are saturated as $\mathsf{SNR}$ increases.

## VI. CONCLUSIONS

In this paper, we investigated multiplexing gain for the two-cell interfering two-user MIMO-MAC using an interference alignment. In addition, by taking the limited feedback system into account, we proposed a feedback framework for interference alignment. On the basis of the proposed feedback framework, we analyzed the rate gap loss and showed that the number of total feedback bits per receiver should be scaled as product of the number of transmit antennas $M$ and $\mathsf{SNR}$ in order to keep a constant rate loss.

## APPENDIX

*Lemma 1*: The beamforming vectors for ICI signal alignment, $\mathbf{v}_1$, $\mathbf{v}_2$, $\mathbf{v}_3$, and $\mathbf{v}_4$, are i.i.d isotropic vectors in $\mathbb{C}^2$.

Proof:) Recall that the beamforming vectors, $\mathbf{v}_1$ and $\mathbf{v}_2$, for ICI signal alignment at $BS_2$, is designed by solving the linear equation as

$$\begin{bmatrix} \mathbf{H}_{2,1} & -\mathbf{H}_{2,2} \end{bmatrix} \begin{bmatrix} \mathbf{v}_1 \\ \mathbf{v}_2 \end{bmatrix} = \mathbf{M}_1 \tilde{\mathbf{v}} = \mathbf{0}. \quad (A1)$$

Since each entry of all channel matrices is i.i.d Gaussian random variable, the concatenation of the beamforming vectors $\tilde{\mathbf{v}}$, which is on the null space of the two concatenated channel matrix, $\mathbf{M}$, is isotropically distributed in $\mathbb{C}^4$. Furthermore, we use the fact that if $\tilde{\mathbf{v}}$ is i.i.d isotropic vector in $\mathbb{C}^4$, the projection of $\tilde{\mathbf{v}}$ onto any subspace is also isotropically distributed in $\mathbb{C}^L$, where $L < 4$. From these facts, we can conclude that $\mathbf{v}_1$, and $\mathbf{v}_2$, are i.i.d isotropic vectors in $\mathbb{C}^2$ because these are the projection of $\tilde{\mathbf{v}}$ in the two orthogonal subspaces. Similarly, $\mathbf{v}_3$, and $\mathbf{v}_4$, are i.i.d isotropic vectors in $\mathbb{C}^2$.

*Lemma 2* (Quantization error of RVQ in [16]): The quantization error between $\hat{\mathbf{v}}_i$ and $\mathbf{v}_i$, $\sin^2\theta_i$, is $2^B\beta(2^B, \frac{M-1}{M})$, where $\beta(\phi,\kappa)$ denotes Beta random variable with parameter $\phi$ and $\kappa$.

*Lemma 3*: The receive combining vector $\mathbf{w}_i$ is isotropically distributed in $\mathbb{C}^3$.

Proof:) Recall that the receive beamforming vector for decoding user 1's message, $\mathbf{w}_1$, is chosen in the nullspace of $[\mathbf{H}_{1,2}\mathbf{v}_2, \mathbf{h}_1^{|\mathsf{CI}|}]$ as in (10). Here, we can argue that the intersection subspace, $\mathbf{h}_1^{|\mathsf{CI}|}$, of two channel matrices $\mathbf{H}_{1,3}$ and $\mathbf{H}_{1,4}$ is isotropically distributed in $\mathbb{C}^3$ because the all entries of the matrices are i.i.d Gaussian random variables. Furthermore, the first column vector $\mathbf{H}_{1,2}\mathbf{v}_2$ is also isotropically distributed in $\mathbb{C}^3$ because it is the projection of the beamforming vector, $\mathbf{v}_2$, with isotropic distribution property (Lemma 1) onto $\mathsf{span}(\mathbf{H}_{1,2})$. Therefore, the receive beamforming vector $\mathbf{w}_1$, which lies on the nullspace of the matrix consisted of these two vectors is isotropically distributed in $\mathbb{C}^3$.

*Lemma 4*: The effective channel direction vector $\bar{\mathbf{g}}_{1,i} = \frac{\mathbf{g}_{1,i}}{\|\mathbf{g}_{1,i}\|}$, $i = \{1,2,3,4\}$, is isotropically distributed in $\mathbb{C}^2$.

Proof:) From the previous definition of effective channel vector after receive combining process, $\frac{\mathbf{g}_{1,i}}{\|\mathbf{g}_{1,i}\|} = \frac{\mathbf{H}_{1,i}^H\mathbf{w}_i}{\|\mathbf{H}_{1,i}^H\mathbf{w}_i\|}$ is the projection of receive combining vector onto $\mathsf{span}(\mathbf{H}_{1,i}^H)$. From lemma 3, $\frac{\mathbf{g}_{1,i}}{\|\mathbf{g}_{1,i}\|}$ is also isotropically distributed in $\mathbb{C}^2$.

## REFERENCES

[1] M. A. Maddah-Ali, A. S. Motahari, and A. K. Khandani, "Communication over MIMO X channels: interference alignment, decomposition, and performance analysis," *IEEE Trans. Inf. Theory*, vol. 54, pp. 3457-3470, Aug. 2008.

[2] S. A. Jafar and S. Shamai, "Degrees of freedom region for the MIMO X channel," *IEEE Trans. Inf. Theory*, vol. 54, pp. 151-170, Jan. 2008.

[3] V. R. Cadambe and S. A. Jafar, "Interference alignment and degrees of freedom of the $K$-User interference channel," *IEEE Trans. Inf. Theory*, vol. 54, pp. 3425-3441, Aug. 2008.

[4] T. Gou and S. A. Jafar "Degrees of freedom for the $K$-user $M \times N$ MIMO interference channel," *Submitted for publication*, arXiv:0809.0099, Aug. 2008.

[5] C. M. Yetis, T. Gou, S. A. Jafar, and A. H. Kayran, "Feasibility conditions for interference alignment," *IEEE Trans. on Signal Processing,* vol. 58, pp. 4771-4782, Sept. 2010.

[6] N. Lee, D. Park, and Y.-D. Kim, "Degrees of freedom on the $K$-user MIMO interference channel with constant channel coefficients for downlink communications,"*in Proc. IEEE Global Communications Conference (GLOBECOM)*, Honolulu, HI, USA, Dec. 2009.

[7] H. Weingarten, S. Shamai, and G. Kramer "On the compound MIMO broadcast channel," *in Proc. IEEE of Annual Information Theory and Applications Workshop* , UCSD, Jan. 2007.

[8] C. Suh and D. Tse, "Interference alignment for cellular networks," *Proc. of Allerton Conference on Communication, Control, and Computing*, Sept. 2008.

[9] W. Shin, N. Lee, J.-B. Lim, C. Shin, and K. Jang, "Interference alignment through user cooperation for two-cell MIMO interfering broadcast channels," *accepted to IEEE Global Communications Conference (GLOBECOM)*, Dec. 2010.

[10] N. Lee, J.-B. Lim, and J. Chun, "Degrees of the freedom of the MIMO Y channel : signal space alignment for network coding," *IEEE Trans. Inf. Theory,* vol. 56, pp. 3332-3342, July 2010.

[11] J. Thukral and H. Bolcskei, "Interference alignment with limited feedback," *in Proc. IEEE ISIT '09*, July 2009.

[12] R. T. Krishnamachari, M. K. Varanasi " Interference alignment under limited feedback for MIMO interference channels," " *Submitted for publication,*, arXiv:0911.5509v1.

[13] R. Tresch, and M. Guillaud, "Cellular interference alignment with imperfect channel knowledge," *in Proc IEEE Int. Conf. on Communications, Workshops*, pp. 1-5, June 2009.

[14] S. Jafar, "Exploiting channel correlations-simple interference alignment schemes with no CSIT ," *arXiv: 0910.0555, 2009.*

[15] S. Jafar, M. Fakhereddin "Degrees of freedom for the MIMO interference channel" *IEEE Trans. Inf. Theory*, vol. 53, No. 7, pp. 2637-2642, July 2007.

[16] C. K. Au-Yeung and Daivd J. Love, "On the performance of random vector quantization limited feedback beamforming in a MISO system," *IEEE Trans. Wireless Communications*, vol. 6, no. 2, pp. 458-462, Feb. 2007.

[17] N. Jindal, "MIMO broadcast channels with finite rate feedback," *IEEE Transactions on Information Theory*, Vol. 52, No. 11, Nov. 2006.

[18] N. Jindal, "Antenna combining for the MIMO downlink channel," *IEEE Trans. Wireless Communications*, vol. 7, no. 10, pp. 3834-3844, Oct. 2008.